\documentclass[aps,prd,twocolumn,groupedaddress,showpacs,nofootinbib,amssymb]{revtex4-1}
\usepackage[T1]{fontenc}
\usepackage[latin1]{inputenc}
\usepackage{graphicx}
\usepackage[english]{babel}
\usepackage{graphicx}
\usepackage{bm}
\usepackage{amsmath}
\usepackage{amssymb}
\usepackage{amsfonts}
\usepackage{epsfig}
\usepackage{colordvi}
\usepackage{color}

\begin{document}
\title{From Clifton-Barrow spherically symmetric to axially symmetric solution in $f(R)$-gravity}
\author{Mariafelicia De Laurentis}

\affiliation{\it Dipartimento di Scienze Fisiche, Università
di Napoli {}``Federico II'', Compl. Univ. di
Monte S. Angelo, Edificio G, Via Cinthia, I-80126, Napoli, Italy,\\
 INFN Sez. di Napoli, Compl. Univ. di
Monte S. Angelo, Edificio G, Via Cinthia, I-80126, Napoli, Italy}
\date{\today}

\begin{abstract}
Starting from Clifton's spherical  static  solution for $f(R)$-gravity we can derive axial symmetric solutions using a methods that takes the advantage of a complex coordinate transformation. The new solutions are generated by applying the Newman-Janis algorithm previously used to transform any static spherically symmetric into axially metric in General Relativity  .
\end{abstract}

\maketitle
\section{Introduction}
\label{uno}
Newman and Janis
showed that it  is possible to  obtain an  axially symmetric  solution (like the
Kerr metric) by making an elementary complex transformation on the
Schwarzschild solution~\cite{pap:nj1}. This same method has been 
used to obtain a new stationary and axially symmetric 
solution known as the
Kerr-Newman metric~\cite{pap:nj2}. The Kerr-Newman space-time is
associated to the exterior geometry of a rotating massive and 
charged black-hole. For a  review on the Newman-Janis method
 to obtain both the Kerr and Kerr-Newman metrics
see~\cite{bk:ier}. 

By means of very elegant
mathematical arguments, Schiffer et al.~\cite{pap:mms} have  given a rigorous
proof to show how the Kerr metric can be derived starting from a complex
transformation on the Schwarzschild solution. We will not go
into the details of this demonstration, but point out  that 
 the proof  relies on two main assumptions. The first  is that the metric
belongs to the same algebraic class of the Kerr-Newman solution,
namely the Kerr-Schild class \cite{pap:gcd}. The second assumption is
that the metric corresponds to an empty solution of the Einstein
field equations. In the case we are going to study, these assumptions are not
considered and hence the proof  in \cite{pap:mms} is not applicable. It is clear, by the
generation of the Kerr-Newman metric, that all the components of
the stress-energy tensor need to be non-zero for the Newman-Janis method to be
successful. In fact, G\"{u}rses and G\"{u}rsey, in
1975~\cite{pap:mg}, showed that if a metric can be written in the Kerr-Schild
form, then a complex transformation ``is allowed in General
Relativity.'' In this paper,  we will show that  such a transformation
can be  extended to $f(R)$-gravity as has already been done by the author in a previous paper \cite{axially}.
 The interest in spherically and axial symmetric solutions of $f(R)$-gravity is growing up because, they could lead to solutions to determine the black holes \cite{odi,lorenzo}. Exact solutions of the field equations are important in order to
gain  insight into the mathematical and physical content of a
theory \cite{bk:dk}, and can be obtained only upon assuming
special  symmetries.
In  \cite{Multamaki},  solutions in vacuum 
have been found considering relations among functions that define the spherical metric or
imposing a costant Ricci curvatue scalar. The authors have reconstructed  the form of some $f(R)$-models,
discussing their physical relevance. In \cite{Multamaki1}, the
same authors have discussed static spherically symmetric  solutions,  in presence of perfect
fluid matter, adopting the metric formalism. They have
shown that a given matter distribution is not capable of globally determining
the functional form of $f(R)$. Others authors have discussed  in
details the spherical symmetry  of $f(R)$-gravity considering also
the relations with the weak field limit.  Exact solutions are
obtained for constant Ricci curvature scalar and for Ricci scalar
depending on the radial coordinate. In particular, it can be considered 
how to obtain results  consistent with GR assuming the well-known post-Newtonian and post-Minkowskian limits as consistency checks.
\cite{spherical}.
In the case of metric $f(R)$ theory {\em in
vacuo}, assuming spherical symmetry does not automatically lead to
the Schwarzschild or Schwarzschild-de Sitter solutions
\cite{spherical}.
The class of
spherically symmetric solutions of metric $f(R)$ gravity is still
unexplored, with only a handful of analytical solutions beyond
Schwarzschild-de Sitter being known \cite{odi,lorenzo}. When one considers axially
symmetric solutions, the situation is even worse: apart from  the
Kerr metric \cite{Kerr}, only the solutions of Ref.~\cite{axially}
are presently available.

 We consider, in particular, a
static spherical solution discovered by Clifton and Barrow
in $f(R)=R^{1+\delta}$ gravity \cite{CliftonBarrow}, which does
not resemble the  Schwarzschild-de Sitter solution, and use it
as a seed metric to generate a new rotating solution which is
not the Kerr metric or a known generalization of it.
The outline of this paper is as follows. In the Sec.\ref{due}, we
introduce  the $f(R)$-gravity action and the field equations with its spherical solutions.
In Sec.\ref{tre}, the Newman-Janis method  is applied to Clifton-Barrow solution.
Discussion and concluding remarks are drawn in Sec. \ref{quattro}.

\section{Field equations and Clifton-Barrow spherically symmetric solutions}
\label{due}
Let us consider a function $f(R)$ of the Ricci scalar $R$ in four dimensions \cite{revi}.
The gravitational action is 

\begin{equation}
{\cal A}=\int
d^4x\sqrt{-g}\biggl[f(R)+\mathcal{X}\mathcal{L}_m\biggr]\,,
\end{equation}
where ${\displaystyle \mathcal{X}=\frac{16\pi G}{c^4}}$ is the coupling,   $\mathcal{L}_m$ is the
standard matter Lagrangian and $g$ is the determinant of the metric\footnote{Here we indicates with "$,$" partial derivative and with " $;$" covariant derivative with regard to  $g_{\mu\nu}$; all Greek indices run from $0,...,3$ and Latin indices run from $1,...,3$; $g$ is the determinant.  }.
The field equations, in metric formalism, read\footnote{All considerations are developed here in metric formalism. From now on we assume physical  units $G=c= 1$.}

\begin{equation}\label{fe1}
f'(R)R_{\mu\nu}-\frac{1}{2}fg_{\mu\nu}-f'(R)_{;\mu\nu}+g_{\mu\nu}\Box_g
f'(R)=\frac{\mathcal{X}}{2}T_{\mu\nu}\,,
\end{equation}
\begin{equation}\label{TrHOEQ}
3\Box f'(R)+f'(R)R-2f(R)\,=\,\frac{\mathcal{X}}{2}T\,,
\end{equation}
with
${\displaystyle T_{\mu\nu}=\frac{-2}{\sqrt{-g}}\frac{\delta(\sqrt{-g}\mathcal{L}_m)}{\delta
g^{\mu\nu}}}$ the energy momentum tensor of matter ($T$ is the
trace), ${\displaystyle f'(R)=\frac{df(R)}{dR}}$ and
$\Box_g={{}_{;\sigma}}^{;\sigma}$. We adopt a $(+,-,-,-)$
signature, while the conventions for Ricci's tensor is
$R_{\mu\nu}={R^\sigma}_{\mu\sigma\nu}$ and 
${R^\alpha}_{\beta\mu\nu}=\Gamma^\alpha_{\beta\nu,\mu}+...$ for the Riemann tensor, where

\begin{equation}\label{chri}
\Gamma^\mu_{\alpha\beta}=\frac{1}{2}g^{\mu\sigma}(g_{\alpha\sigma,\beta}+g_{\beta\sigma,\alpha}-g_{\alpha\beta,\sigma})\,,
\end{equation}
are the Christoffel symbols  of the $g_{\mu\nu}$
metric.
We choice $f(R) = R^{1+\delta}$ which reduces to General Relativity in the limit $\delta\rightarrow0$, and we consider only the exact static spherically symmetric vacuum solutions deriving from this Lagrangian  \cite{CliftonBarrow}. The static solution is given by the line elements
\begin{equation}
ds^2=A(r)dt^2-\frac{dr^2}{B(r)}-r^2d\Omega^2\,,
\label{first}
\end{equation}
where
\begin{eqnarray*}
A(r)&=& r^{2\delta \frac{1+2\delta}{1-\delta}}+\frac{C}{r^{\frac{1-4\delta}{1-\delta}}}\,,\nonumber\\
B(r)&=&\frac{(1-\delta)^2}{(1-2\delta+4\delta^2)(1-2\delta(1+\delta))}\left(1+\frac{C}{r^{\frac{1-2\delta+4\delta^2}{1-\delta}}}\right)\,,\nonumber\\
\label{AB1}
\end{eqnarray*}
and $C$ is a constant. 
The task is now to show how, from  a spherically symmetric solution, one can  generate an axially symmetric solution adopting the Newman-Janis procedure that works in General Relativity. 
%
%
\section{Axial symmetry for $f(R)=R^{1+\delta}$ gravity}
\label{tre}

We want to show, now, how it is possible to obtain an axially symmetric solution starting from a spherically symmetric  seed solution (\ref{first}) adopting the method developed by Newman and Janis in General Relativity \cite{pap:nj1,axially}. Such an algorithm can be applied to a
static spherically symmetric  metric, considered as a ``seed'' metric. 
Following Newman and Janis,
Eq.~(\ref{first}) can be written in the so called 
Eddington--Finkelstein coordinates $(u,r,\theta,\varphi)$, {\it i.e.} the
$g_{rr}$ component is eliminated by a change of coordinates and a
cross term is introduced \cite{gravitation}. Specifically this is achieved by defining
the time coordinate as $dt = du + F(r)dr$ and setting 

\begin{equation}
F(r)= \left[\frac{A(r)}{B(r)}\right]^{-\frac{1}{2}}\,
=\,\frac{r^{2\delta-1}}{\sqrt{\Psi}\left(C+r^{\frac{1-2\delta+4\delta^2}{
1-\delta}}\right)}\,,
\end{equation}
where for simplicity we set $\Psi=\frac{(1-\delta)^2}{(1-2\delta+4\delta^2)[1-2\delta(1+\delta)]}$
Once such a transformation is performed,  the metric
(\ref{first})  becomes
\begin{eqnarray}\label{nullelemn}
 ds^2 &=& A(r) \, du^2 + 2 \sqrt{A(r)B(r)} \, du \, dr -
r^2d\Omega^2 \,=\nonumber\\ &&=\,A(r)du^2 +
2 r^{\frac{\delta(2\delta+1)}{1-\delta}}\left(\Psi \right)^{-\frac{1}{2}}
\, du \, dr -
r^2d\Omega^2 \,.\nonumber\\
\end{eqnarray}
The surface $u\,=\,$ costant is a light cone starting 
from the origin $r\,=\,0$.
The metric tensor for the line element
(\ref{nullelemn}) in controvariant form and null-coordinates is

\begin{eqnarray}
ds^2 &=&2\left(\Psi\right)^{-\frac{1}{2}}r^{\frac{\delta(2\delta+1}{1-\delta}}dudr-\nonumber\\ &&+\Psi \left[1+C \left(r^\frac{1-2\delta+4\delta^2}{\delta-1}\right)^{-1}\right]dr^2-\frac{1}{r^2}d\theta^2-\nonumber\\&&+\frac{1}{r^2\sin^2\theta}d\varphi^2\nonumber\\
\label{contrometric}
\end{eqnarray}

The metric  (\ref{contrometric}) can be written in terms of a null
tetrad as
\begin{equation}\label{eq:gmet}
g^{\mu\nu} = l^\mu n^\nu + l^\nu n^\mu-m^\mu\bar{m}^\nu -
m^\nu\bar{m}^\mu,
\end{equation}
where $l^\mu$, $n^\mu$, $m^\mu$ and $\bar{m}^\mu$ are the vectors
satisfying the conditions

\begin{eqnarray}&& l_\mu l^\mu\,=\,m_\mu m^\mu\,=\,n_\mu n^\mu\,=\,0,\nonumber\\
\\
&&l_\mu n^\mu\,=\,-m_\mu\bar{m}^\mu\,=\,1, \nonumber\\
\\
&&l_\mu m^\mu\,=n_\mu m^\mu\,=\,0\,.\nonumber\\ \end{eqnarray} The bar indicates the
complex conjugation. At any point in space, the tetrad can be chosen
in the following manner: $l^\mu$ is the outward null vector
tangent to the cone, $n^\mu$ is the inward null vector pointing
toward the origin, and $m^\mu$ and $\bar{m}^\mu$ are the vectors
tangent to the two-dimensional sphere defined by constant $r$ and
$u$ (see \cite{chandra}).   For the space-time (\ref{contrometric}), the tetrad null
vectors can be

\begin{equation}
\left\{\begin{array}{ll}
l^\mu =  \delta^\mu_1 \,,\\
\\
n^\mu =  -\frac12\left[\Psi \left(1+\frac{C}{r^\frac{1-2\delta
+4\delta^2}{\delta-1}}\right) \right]\delta^\mu_1+
\frac{\sqrt{\Psi}}{r^{\frac{\delta \left( 2\delta+1
\right)}{1-\delta}}} \, \delta^\mu_0 \,, \\
\\
m^\mu  =  \frac{1}{\sqrt{2}r} \left( \delta^\mu_2
+\frac{i}{\sin{\theta}} \, \delta^\mu_3 \right) \,, \\
\\
\bar{m}^\mu  =  \frac{1}{\sqrt{2}r} \left( \delta^\mu_2
-\frac{i}{\sin{\theta}}\delta^\mu_3 \right)\,.\end{array}\right. \,.
\end{equation}

Now we need to extend the set of coordinates
$x^\mu\,=\,(u,r,\theta,\varphi)$ replacing the real
radial coordinate by a complex variable. Then the tetrad
null vectors become \footnote{It is worth noticing that a certain
arbitrariness  is present in the 
complexification process of the functions.  Obviously, we have to obtain the metric (\ref{contrometric})  as soon as $r\,=\,\bar{r}$.}

\begin{equation}\label{tetradvectors}
\left\{\begin{array}{ll}
l^\mu = \delta^\mu_1 \,,\\
\\
n^\mu =  -\frac{1}{2} \Psi \left[
\frac{(r\bar{r})^{\beta}+C
(r+\bar{r})^{\beta}}{(r\bar{r})^{\beta}}\right] \delta^\mu_1+ \frac{\sqrt{\Psi}}{(r
\bar{r})^{\frac{\delta(2\delta+1)}{2(1-\delta)}}}
\delta^\mu_0 \,,\\
\\
m^\mu =  \frac{1}{\sqrt{2}\bar{r}} \left( \delta^\mu_2
+\frac{i}{\sin{\theta}}\delta^\mu_3 \right)\,,\\
\\
\bar{m}^\mu  =  \frac{1}{\sqrt{2}r} \left( \delta^\mu_2
-\frac{i}{\sin{\theta}}\delta^\mu_3 \right)\,.
\end{array}\right.
\end{equation}
where $\beta=\frac{1-2\delta+4\delta^2}{\delta-1}$.
A new metric is obtained by making a complex coordinates
transformation

\begin{equation}\label{transfo}x^\mu\rightarrow\tilde{x}^\mu=x^\mu+iy^\mu(x^\sigma)\,,\end{equation}
where $y^\mu(x^\sigma)$ are analityc functions of the real
coordinates $x^\sigma$, and simultaneously let the null 
tetrad vectors $Z^\mu_a\,=\,(l^\mu,n^\mu,m^\mu,\bar{m}^\mu)$, with
$a\,=\,1,2,3,4$, undergo the transformation

\begin{equation}Z^\mu_a\rightarrow \tilde{Z}^\mu_a(\tilde{x}^\sigma,\bar{\tilde{x}}^\sigma)\,=\,Z^\rho_a\frac{\partial\tilde{x}
^\mu}{\partial x^\rho}.\end{equation} Obviously, one has to recover the old tetrads and metric as soon as 
$\tilde{x}^\sigma\,=\,\bar{\tilde{x}}^\sigma$. In summary, the
effect of the "\emph{tilde transformation}" (\ref{transfo}) is to
generate a new metric whose components are (real) functions of
complex variables, that is

\begin{equation}g_{\mu\nu}\rightarrow \tilde{g}_{\mu\nu}\,:\,\tilde{\mathbf{x}}\times\tilde{\mathbf{x}}\mapsto
\mathbb{R}\end{equation} with

\begin{equation}\tilde{Z}^\mu_a(\tilde{x}^\sigma,\bar{\tilde{x}}^\sigma)|_{\mathbf{x}=\tilde{\mathbf{x}}}
=Z^\mu_a(x^\sigma).\end{equation} For our aims,  we can make
 the choice

\begin{equation}\label{transfo_1}
\tilde{x}^\mu\,=\,x^\mu+ia(\delta^\mu_1-\delta^\mu_0)\cos\theta\rightarrow
\left\{\begin{array}{ll}\tilde{u}\,=\,u+ia\cos\theta\\\\\tilde{r}\,=\,r-ia\cos\theta\\\\
\tilde{\theta}\,=\,\theta\\\\
\tilde{\phi}\,=\,\phi\\\\
\end{array}\right.\end{equation}
where $a$ is constant and the tetrad null vectors
(\ref{tetradvectors}), if we choose
$\tilde{r}\,=\,\bar{\tilde{r}}$, become

\begin{equation}\label{tetradvectors_2}
\left\{\begin{array}{ll}\tilde{l}^\mu\,=\,\delta^\mu_1\\
\\
\tilde{n}^\mu=-\frac{1}{2} \Psi \left\{ 1+C [
(\Sigma^{-2})\Re(\tilde{r})]^{\beta}\right\}
\delta^\mu_1+\,\sqrt{\Psi}\,
\Sigma\,^{\frac{\delta(2\delta+1)}{(\delta-1)}}\,
\delta^\mu_0 \,, \\
\\
\tilde{m}^\mu = 
\frac{1}{\sqrt{2}(\tilde{r}
-ia\cos\theta)}\biggl[ia(\delta^\mu_0-\delta^\mu_1)
\sin\theta+\delta^\mu_2
+\frac{i}{\sin{\theta}} \, \delta^\mu_3\biggr] \,,\\
\\
\bar{\tilde{m}}^\mu  =   \frac{1}{\sqrt{2}(\tilde{r}
+ia\cos\theta)}\biggl[-ia(\delta^\mu_0-\delta^\mu_1)
\sin\theta+\delta^\mu_2
-\frac{i}{\sin{\theta}} \, \delta^\mu_3\biggr] \,,
\end{array}\right.
\end{equation}
where $\Sigma = \sqrt{\tilde{r}^2 + a^2\cos^2{\theta}}$.

From the transformed null tetrad vectors, a new metric is recovered
using (\ref{eq:gmet}). For the null tetrad vectors given by
(\ref{tetradvectors_2}) and the transformation given by
(\ref{transfo_1}), the new metric, with coordinates
$\tilde{x}_\mu\,=\,(\tilde{u},\tilde{r},\theta,\varphi)$, is
\begin{eqnarray}
ds^2 &=&\Sigma^ {\frac{2\delta
\left( 2 \delta+1 \right)}{1-\delta}} \left[ 1+ C
\left(\frac{\Re(\tilde{r})}{\Sigma^{2}} \right)^{\beta} \right]d\tilde{u}^2+\nonumber\\&&+2\frac{\Sigma^{\frac{\delta(2
\delta+1)}{1-\delta}}}{\sqrt{\Psi}}d\tilde{u}\tilde{r}+2a \Sigma^{\frac{2\delta
\left( 2 \delta+1 \right)}{1-\delta}}\times\nonumber\\&&\times \left[ \sqrt{\Psi}
\Sigma^{\frac{\delta \left( 2 \delta+1 \right)}{\delta-1}}- 1
-C \left(
\frac{\Re(\tilde{r})}{\Sigma^{2}}\right)^{\beta}
\right] \times\nonumber\\&&\times\sin^2 \theta d\tilde{u} \tilde{\varphi}-
2a \frac{\Sigma^{\frac{\delta(1+2\delta)}{1
-\delta}}\sin^2{\theta}}{\sqrt{\Psi}} 
 d \tilde{r} \tilde{\varphi}- \Sigma^2d\tilde{\theta}^2-\nonumber\\&&-\left\{\Sigma^2 +
a^2\sin^2{\theta} \Sigma^{\frac{2\delta \left( 2 \delta+1
\right)}{1-\delta}}
\left[ \sqrt{\Psi}\Sigma^{\frac{\delta(2
\delta+1)}{\delta-1}}-\right.\right.\nonumber\\&&+\left.\left. 1-C \left(
\frac{\Re(\tilde{r})}{\Sigma^{2}}\right)^{\beta} \right]\right\}\sin^2{\theta}d\tilde{\varphi}^2
\label{eqn:coform}
\end{eqnarray}
The form of this metric gives the
general result of the Newman-Janis algorithm starting from  any spherically
symmetric "seed" metric.
The metric given in Eq.~(\ref{eqn:coform}) can be  simplified by a further gauge  transformation so that the only off-diagonal
component is $g_{\varphi t}$. This procedure makes it easier to compare with
the standard Boyer-Lindquist form of the Kerr metric
\cite{gravitation} and to interpret physical properties such as the frame
dragging. The coordinates $\tilde{u}$ and $\varphi$ can be
redefined in such a way that the metric in the new coordinate
system has the properties described above. More explicitly, if we
define the coordinates in the following way
\begin{equation}d\tilde{u}\,=
\,dt+g(\tilde{r})d\tilde{r}\,\,\,\,\,\text{and}\,\,\,\,\,d\varphi\,=\,d\varphi+h(\tilde{r})d\tilde{r}\end{equation}
where
\begin{equation}
\left\{\begin{array}{ll}
g(\tilde{r})= -\frac{ \sqrt{\Psi} \,
\Sigma^{\frac{2 \delta^2+ 3 \delta
-2}{\delta-1}}\,+\,a^2\sin^2{\theta} }{\Psi \Sigma^2
\left[ 1 + C  \left(
\frac{\Re\left\{\tilde{r}\right\}}{\Sigma^{2}}
\right)^{\beta} \right]  +
a^2\sin^2{\theta} }   \\
\\
h(\tilde{r})
=-\frac{ a} {\Psi \Sigma^2 \left[ 1 + C
 \left(  \frac{\Re\left\{ \tilde{r}\right\}}{\Sigma^{2}}
\right)^{\beta} \right]  +
a^2\sin^2{\theta} } 
\end{array}\right.
\end{equation}
after some algebraic manipulations, one finds that, in
$(t,\tilde{r},\theta,\varphi)$ coordinates system, the metric
(\ref{eqn:coform}) becomes
\begin{eqnarray}
\label{eq:blform}
ds^2&=& \Sigma^{\frac{2\delta(2 \delta+1)}{1-\delta}}
\left[ 1+ C \left(\frac{\Re(\tilde{r})}{\Sigma^{2}}
\right)^{\beta} \right]dt^2+\nonumber\\&&+2 a \Sigma^{\frac{2\delta(2
\delta+1)}{1-\delta}}
\left[ \sqrt{\Psi} \Sigma^{\frac{\delta(2
\delta+1)}{\delta-1}}-  1-C\left(
\,\frac{\Re(\tilde{r})}{\Sigma^{2}}\right)^{
\beta}
\right] \times\nonumber\\ &&\times
\sin^2{\theta}dtd\varphi-
\left(\Psi\left[1+C\left(\frac{\Re\{\tilde{r}\}}{\Sigma^2}\right)^{
\beta}\right]+\right.\nonumber\\&&+\left.
a^2\sin^2{\theta}\right)^{-1}dr^2- \Sigma^2 d\theta^2
-\Biggl\{\Sigma^2 +\Biggr.\nonumber\\&&\Biggl.+ a^2
\Sigma^{\frac{2\delta(2 \delta+1)}{1-\delta}} \sin^2{\theta}
\left[\sqrt{\Psi} \Sigma^{\frac{\delta(2
\delta+1)}{\delta-1}}-\Biggr.\right. \nonumber\\&&\left.\Biggr.+ 1
\,-\, C \, \left(
\,\frac{\Re(\tilde{r})}{\Sigma^{2}}\right)^{\beta}
\right]\Biggr\}\sin^2{\theta}d\varphi^2
\end{eqnarray}
This metric is the product of the  Newman-Janis algorithm
applied to the Clifton-Barrow static spherically symmetric
solution of $ R^{1+\delta}$ gravity,  in
Boyer-Lindquist type coordinates. In the same manner,
axisymmetric solutions of other  $f(R)$ gravity theories can
be generated. This metric describes a 2-parameter family of
solutions with the parametr $C$ corresponding to the mass of
the central object and $a$ to the angular momentum per unit
mass.  Our result generalizes those previously obtained in
Ref.~\cite{axially} since, in the special case
$ \delta = 1/4 $, not only one obtains immediately
the spherically symmetric space-time
\begin{equation}\label{sol_noe}
ds^2= \left( C+ r \right) dt^2 - \frac{1}{2}\frac{ r}{C+ r}
dr^2 - r^2 d\Omega^2
\end{equation}
but, in addition, the  axial metric  (\ref{eq:blform}) reduces 
to the spherycally symmetric metric found
in \cite{axially}.

\section{Conclusions}
\label{quattro}

The Newman-Janis method was developed for General Relativity but,
as shown in Sec.\ref{tre}, it can be applied successfully also to metric
$f(R)$-gravity. The Clifton-Barrow static and spherically
symmetric solution is mapped into the new stationary axisymmetric
solution (\ref{eq:blform}), which is the main
result of this paper. Due to the fourth order of its field
equations, metric $f(R)$-gravity has a richer variety of solutions
than General Relativity and the Kerr solution is not the only
vacuum solution of this theory, nor the most general. The new
metric testifies of this fact. In addition, it does not belong to
the class of axisymmetric solutions discovered in \cite{axially}.

\section{Acknowledgement}
I acknowledge Prof. Salvatore Capozziello for fruitful discussions, comments and suggestions on this work.


\end{document}